# Spectroscopic properties of Cr,Yb:YAG nanocrystals under intense NIR radiation

M. Chaika, R. Tomala, O. Bezkrovnyi, W. Strek

Institute of Low Temperature and Structure Research, Polish Academy of Sciences, Wroclaw, Poland

**Abstract**

Laser induced white emission (LIWE) was thoroughly studied in recent decades. However, despite the progress in understanding of this phenomenon, the mechanism behind LIWE remains unclear. The present paper focuses on the influence of Yb content on the LIWE properties of Cr,Yb:YAG nanocrystals. Microstructure and optical properties of the samples were characterized and the influence of the concentration of $Yb^{3+}$ ions on the spectroscopic properties of Cr,Yb:YAG and energy transfer processes between $Cr^{3+}$ and $Yb^{3+}$ ions was revealed. Multiphoton ionization theory was used to explain the findings of the paper.

## 1. Introduction

Since the discovery of laser induced white emission (LIWE) in 2010 by Wang and Tanner, its understanding was greatly improved, but no complete model was proposed. This phenomenon is based on the generation of strong emission covering entire visible and part of the near-infrared region under focused laser beam in vacuum. LIWE is promising for use in artificial light sources, converters for solar panels, nanoheaters, etc. [1,2]. Moreover, this phenomenon is already used for hydrogen generation [3]. However, the nature of LIWE phenomenon remains unclear. So far, various theories have been proposed to explain LIWE, but there is no complete theory that can explain all the features of this phenomenon. LIWE can be generated in different types of materials from inorganic phosphors to organometallic nanostructures. Despite the differences in the structures of the studied materials, LIWE properties remain the same including broadband emission spectra, pressure dependence, power dependence, etc. So, it can be concluded that the same mechanism must be responsible for LIWE in all these materials. However, the explanations proposed up to now are based on the specific properties of the studied materials and often fail to explain all the features of LIWE.

The proposed models provide explanation of individual features of LIWE in specific materials and often fail to describe the whole picture. The main feature of LIWE is the strong broadband emission in the visible region, which is produced when the sample is placed in a vacuum environment under laser excitation. In fact, white light emission can occur in various atmospheres [4]. An increase in pressure in the measuring chamber causes a decrease in the emission intensity. Authors associated this feature with a thermal avalanche process when increase in pressure leads to fast disappearance of thermal energy. However, in some cases, increase in the temperature of the sample caused a decrease in LIWE intensity [5] which contradicts the proposed theory. On the other hand, a threshold behavior and an exponential law of power dependence of LIWE intensity indicate that an avalanche-like process is involved in this phenomenon. Intervalence charge transfer model successfully explains the concentration dependence of LIWE in Re-doped phosphors and the fact that Yb-doped materials show higher emission intensity than other materials. Models of electron-hole recombination, excitation of cage oxygen, $sp^2$-$sp^3$ recombination, defect luminescence, etc. [2] are related to specific materials and cannot be adapted to other ones.

Based on our earlier research, it was concluded that a modified multiphoton ionization model can be proposed to describe the nature of LIWE. Our recent discoveries have shown that LIWE can be generated only at the surface of the sample and do not penetrate into the volume [6]. White light emission was obtained at the entrance and/or at the exit of the laser beam without any generation inside the volume [5]. Using this discovery multiphoton ionization was proposed to be a key part of LIWE process [1]. Previously, using multiphoton ionization the interaction of femtosecond laser pulses of high density with atoms accompanied by formation of photoelectrons was explained. This model is based on the ionization of atoms via simultaneous absorption of several photons. The application of this model is based on the similarity between LIWE and multiphoton ionization (power threshold, exponential rise in emission intensity, saturation, etc.). This model already was used to explain some interesting features of LIWE, such as white light pulsation [1], and memory effect [7]. Despite the progress in understanding of LIWE, the current model is far from perfect.

The aim of the present paper is to improve the current knowledge of LIWE phenomenon by studying the spectroscopic properties of Cr,Yb:YAG nanocrystals under an influence of NIR excitation. The influence of energy transfer between $Cr^{3+}$ and $Yb^{3+}$ ions on LIWE properties is shown. Multiphoton ionization model is used to explain the findings of the paper.

## 2. Experimental

Cr,Yb:YAG nanocrystals were synthetized by Pechini method. Seven samples with different concentrations of Yb ions (1%, 5%, 10%, 30%, 30%, 50%, 70% and 100%) relative to Y ions were prepared. Except for two samples (pure YAG and Yb:YAG), the concentrations of Cr ions were the same - 0.5 at.% relative to Al ions. The list of synthesized samples is collected in the Table 1.

Table 1: Chemical compositions of synthesized Cr,Yb:YAG nanocrystals.

| Denote | Yb0* | Yb1* | Yb0 | Yb1 | Yb5 | Yb10 | Yb20 | Yb30 | Yb50 | Yb70 | Yb100 |
|---|---|---|---|---|---|---|---|---|---|---|---|
| Yb, at.% | 0 | 1 | 0 | 1 | 5 | 10 | 20 | 30 | 50 | 70 | 100 |

| Cr, at.% | 0 | 0 | 0.5 | 0.5 | 0.5 | 0.5 | 0.5 | 0.5 | 0.5 | 0.5 | 0.5 |

The microstructure, optical, and LIWE properties of synthetized Cr,Yb:YAG nanocrystals were investigated. The X-ray diffraction patterns were measured using Panalytical X'Pert pro X-ray powder diffractometer. The synthesized samples were analyzed by transmission electron microscopy using the Philips CM-20 SuperTwin microscope. Absorbance spectra were measured by a Varian 5E UV-VIS-NIR spectrophotometer. Excitation / emission spectra and lifetimes were obtained at room temperature using an Edinburgh Instruments FLS980 fluorescence spectrometer. 975 nm laser was used as the excitation source. LIWE spectra were collected in vacuum using an AVS-USB2000 Avantes and Ocean Optics NIRQuest512-2.5 spectrometers for the anti-Stokes and Stokes parts of the spectra. One can find a detailed explanation of the experimental part in supplementary files.

## 3. Result and Discussion

### 3.1 Microstructure

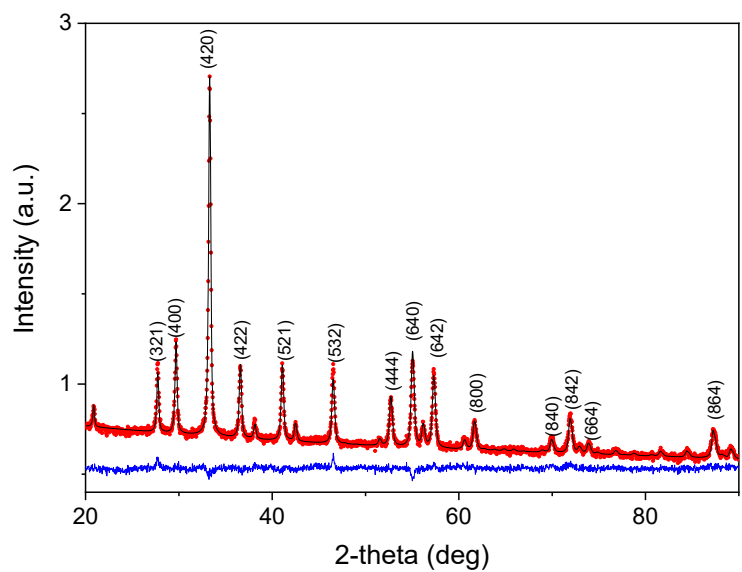

Fig. 1: X-ray diffraction pattern of the 1 at.% doped Cr,Yb:YAG nanocrystals (black curve), results of Rietveld refinement analysis (red and blue curve).

The microstructure of the samples was studied by the XRD techniques. X-ray diffraction studies revealed a pure YAG phase, the average grain size was ~ 30 nm for all samples. Fig 1 shown X-ray diffraction pattern of the Yb1 sample, other patterns are shown in supplementary files (Fig. S1). XRD patterns consist of the main peaks belonging to Ia3d space group related to the garnet phase.

No extra peaks related to other phases were detected in the samples. Diffraction data were analyzed by Rietveld analysis. Table 2 summarizes the results of XRD analysis including cell parameters, crystallite sizes, and lattice microstrain. The calculated lattice parameters are in the range from 11.9 Å to 12.3 Å. Increase in Yb concentration in Cr,Yb:YAG nanocrystals leads to decrease in the lattice parameters according to Vegard's law similar to the results reported earlier [8]. Average grain sizes and lattice microstrain range from 25 to 35 nm, and from 0.0016 to 0.0023, respectively.

Table 2: Crystallographic parameters of the garnet phase of Cr,Yb:YAG nanocrystals obtained by the Rietveld method: Cell parameter a, crystallite size $D_{XRD}$, lattice microstrain $\varepsilon$.

| Yb, at.% | Cr, at.% | a, Å | $D_{XRD}$, nm | $\varepsilon$ |
|---|---|---|---|---|
| 0 | 0 | 12.0125(6) | 25(1) | 0.0023(2) |
| 0 | 5 | 12.0312(5) | 34(1) | 0.0017(1) |
| 1 | 0 | 12.0229(9) | 27(1) | 0.0021(1) |
| 1 | 5 | 12.0237(9) | 35(1) | 0.0016(1) |
| 5 | 5 | 12.0202(8) | 35(1) | 0.0016(1) |
| 10 | 5 | 12.0147(9) | 36(1) | 0.0016(1) |
| 20 | 5 | 12.0142(9) | 33(1) | 0.0017(1) |
| 30 | 5 | 12.0088(9) | 32(1) | 0.0018(1) |
| 50 | 5 | 11.9954(9) | 31(1) | 0.0018(1) |
| 70 | 5 | 11.9801(9) | 30(1) | 0.0019(1) |
| 100 | 5 | 11.9594(9) | 33(2) | 0.0017(1) |

TEM reveal a fine crystal structure of synthesized Cr,Yb:YAG nanocrystals confirming thereby the results of the XRD studies. TEM image of the Yb5 sample is shown in the Fig. 2. The samples are characterized by the fine crystalline structure, no traces of an amorphous phase were found on the surfaces of the samples (inset in the Fig. 2). All studied samples have the same morphology; the average particle sizes are in the range of several tens of nanometers, which is consistent with the XRD data.

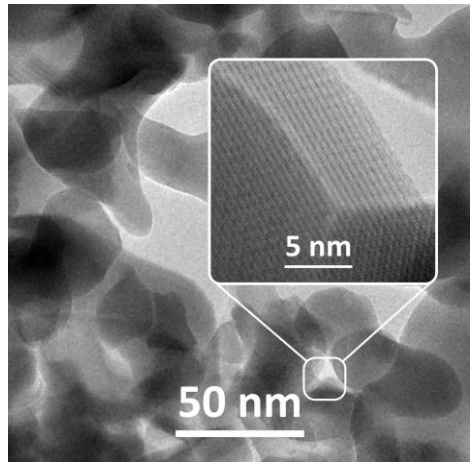

Fig. 2: TEM image of Cr,Yb:YAG nanocrystals.

## 3.2 Optical properties

Diffuse reflectance spectra of Cr,Yb:YAG nanocrystals show the presence of absorption bands corresponding to $Yb^{3+}$, $Cr^{3+}$, $Cr^{6+}$ ions, and color centers. Fig. 3 shows diffuse reflectance spectra of the samples. The spectra are characterized by the presence of narrow absorption peaks in the range from 850 to 1100 nm corresponding to $^2F_{7/2} \rightarrow {}^2F_{5/2}$ electronic transitions of $Yb^{3+}$ ions [9,10]. An increase in the concentration of ytterbium ions leads to an increase in the intensity of $Yb^{3+}$ absorption bands. The absorption bands originated from $Yb^{3+}$ ions are centered at 915 nm, 927 nm, 939 nm, 969 nm, and 1030 nm. The main absorption band centered at 969 nm is similar to the previously reported one for Yb:YAG nanocrystals [8] being blue shifted by 6 nm compared to the single crystal [9]. Two absorption bands centered at 450 nm and 600 nm correspond to the $^4A_{2g} \rightarrow {}^4T_{2g}$ and $^4A_{2g} \rightarrow {}^4T_{2g}$ transitions of $Cr^{3+}$ ions, respectively [11–15]. No $Yb^{2+}$ absorption bands were observed. It should be noted that $Yb^{2+}$ ions are characterized by strong oscillator strengths, so even a small amount of $Yb^{2+}$ ions should be visible in diffuse reflectance spectra [10].

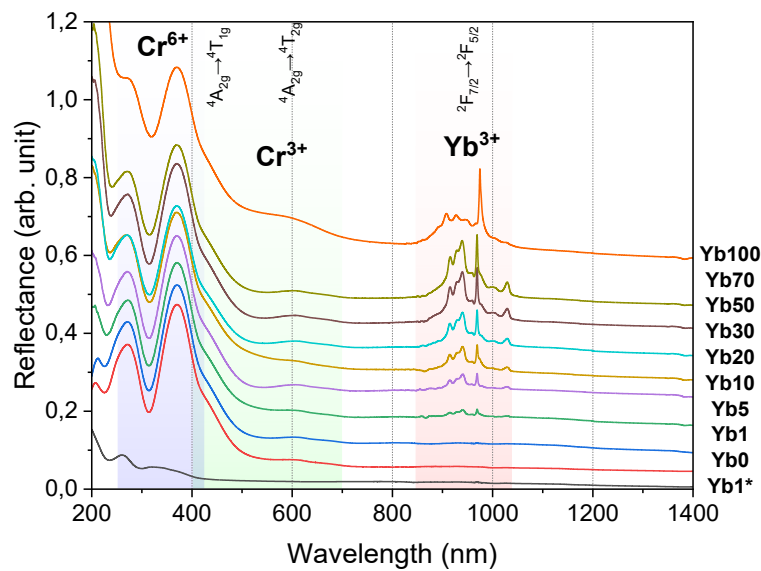

Fig. 3: Diffuse reflectance spectra of Cr,Yb:YAG nanocrystals.

Doping by chromium ions of Yb:YAG nanocrystals caused an increase in absorption in the UV region, probably due to appearance of $Cr^{6+}$ ions. The diffuse reflectance spectra of YAG nanocrystals doped only by $Yb^{3+}$ (1 at.%) are characterized by the presence of absorption bands at 210 nm, 260 nm, and multiple bands at ~330 nm (Fig. 3, Yb1*). The possible origin of the band centered at 260 nm is a charge transfer transition between $Fe^{3+}$ ions oxygen ligands [9], while the bands centered at 330 nm most likely originate from color centers. However, YAG nanocrystals doped only with chromium (0.5 at.%) are characterized by the presence of additional strong absorption bands at 270 nm and 370 nm, besides the main $Cr^{3+}$ absorption bands centered at 430 nm and 600 nm. As can be seen, an addition of ytterbium in concentrations up to 100 at.% does not change the position and absorption intensity of these additional bands, which indicates that they are related to chromium ions. Most likely, strong absorption bands centered at 270 nm and 370 nm correspond to the charge transfer transitions from oxygen to $Cr^{6+}$ ions [16]. It should be noted that an increase in the concentration of ytterbium leads to an increase in the absorption near 210 nm, which is most likely due to an increase in the amount of defects (first of all, oxygen vacancies) [17]. However, the origin of the absorption centers at ~210 nm remains unclear. This absorption can be related to the band gap transition (VB→CB).

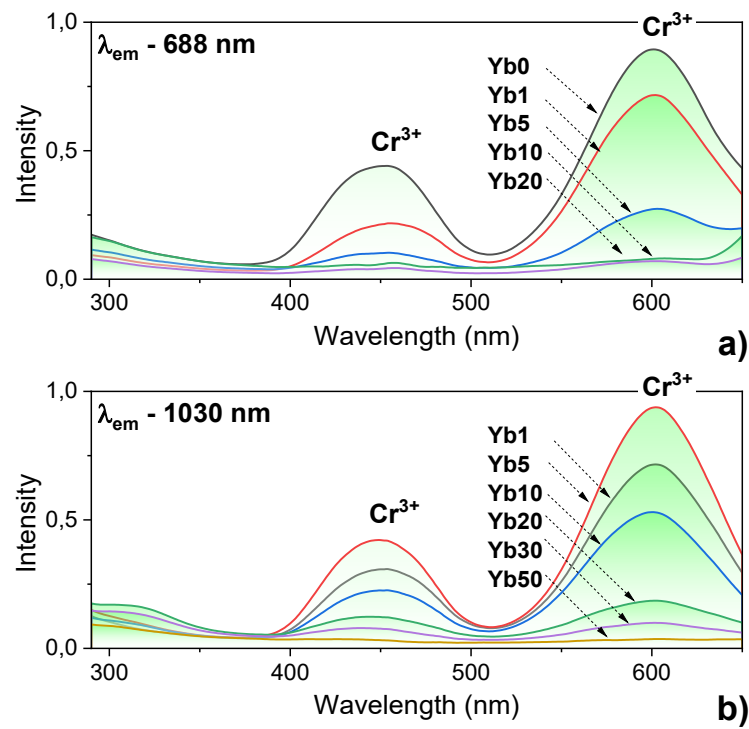

Fig. 4: Excitation spectra of Cr,Yb:YAG nanocrystals measured at a) $\lambda_{em}$ = 688 nm and $\lambda_{em}$ = 1030 nm at room temperature.

The excitation spectra of $Cr^{3+}$ and $Yb^{3+}$ ions follow a similar pattern indicating the energy transfer process between these ions. Excitation spectra of Cr,Yb:YAG nanocrystals were investigated (Fig. 4). The measured spectra were recorded at $\lambda_{em}$ = 688 nm and 1030 nm, which correspond to $^4T_{2g} \rightarrow {}^4A_{2g}$ and $^2E_g \rightarrow {}^4A_{2g}$ transitions of $Cr^{3+}$ ions, and $^2F_{5/2} \rightarrow {}^2F_{7/2}$ transition of $Yb^{3+}$ ions, respectively [18,19]. Two broad bands corresponding to $^4A_{2g} \rightarrow {}^4T_{2g}$ and $^4A_{2g} \rightarrow {}^4T_{2g}$ transitions of $Cr^{3+}$ ions are observed in the excitation spectra. No bands corresponding to $Cr^{6+}$ ions were detected (Fig. 4(a)) indicating that in our case the charge transfer process between $Cr^{6+}$ and $Cr^{3+}$ ions either does not occur or is weak. The excitation spectra of $Yb^{3+}$ ions show a similar pattern to that of $Cr^{3+}$ ions (Fig. 4(b)), indicating that energy transfer occurs between $Cr^{3+}$ and $Yb^{3+}$ ions. It should be noted that both excitation spectra of $Cr^{3+}$ and $Yb^{3+}$ ions show emission upon UV excitation. An additional band centered at ~330 nm can be caused by the $^4A_2 \rightarrow {}^4T_1(te^2)$ transition in $Cr^{3+}$ ions.

The photoluminescence spectra of $Cr^{3+}$ ions show a decrease in total emission intensity with increasing $Yb^{3+}$ concentration due to energy transfer from $Cr^{3+}$ ions to $Yb^{3+}$ ones. Fig. 5 shows the emission spectra of $Cr^{3+}$ ions in Cr,Yb:YAG nanocrystals under 450 nm excitation. The spectra consist of a narrow $^2E_g \rightarrow {}^4A_{2g}$ R-line (688 nm) together with a diffuse band centered at 675 nm (anti-Stokes vibronic sidebands), and two sharp diffuse bands centered at 703 nm and 725 nm corresponding to vibronic sidebands accompanying the fluorescence transition [14]. Broadband $Cr^{3+}$ emission band at 707 nm is attributed to $^4T_{2g} \rightarrow {}^4A_{2g}$ transitions. The spectra show a decrease in the emission intensity of $Cr^{3+}$ ions with an increase in $Yb^{3+}$ concentration. Compared with the $Cr^{3+}$ single-doped sample, the $Cr^{3+}$ photoluminescence intensity gradually decreases from 73% for the sample doped with 1 at.% of $Yb^{3+}$ ions to 0,6% for 100% $Yb^{3+}$-doped sample. The change in the luminescence intensity for Cr,Yb:YAG samples is shown in Table 3. The decrease in the emission intensity occurs due to energy transfer from $Cr^{3+}$ ions to $Yb^{3+}$ ions, as increasing the concentration of $Yb^{3+}$ ions causes an increase in energy transfer efficiency.

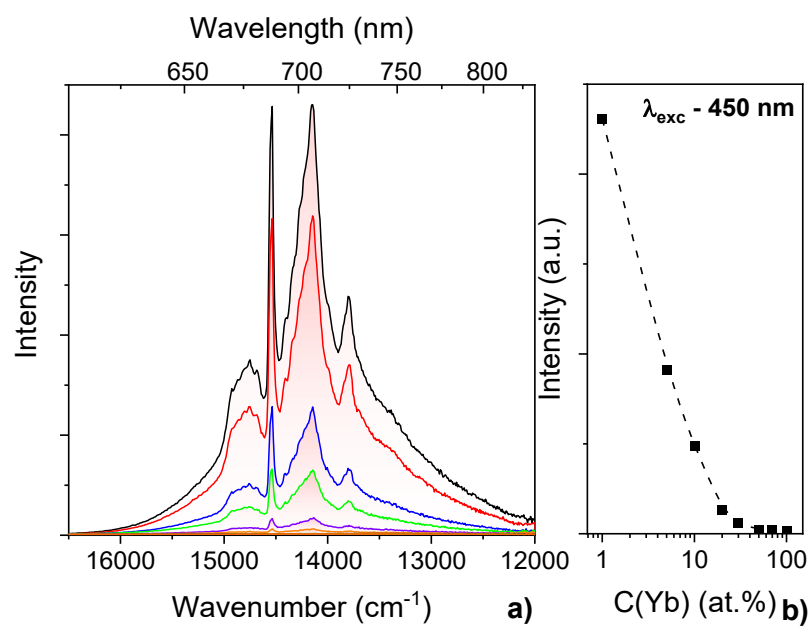

Fig. 5: a) Photoluminescence spectra of $Cr^{3+}$ ions. b) Dependence of luminescence intensity (450 nm) on the concentration of $Yb^{3+}$ ions in Cr,Yb:YAG nanocrystals at room temperature.

Table 3: Relative emission intensities of $Cr^{3+}$ and $Yb^{3+}$ ions in Cr,Yb:YAG nanocrystals: $Cr^{3+}$ emission intensity at 450 nm ($I_{450}$), $Yb^{3+}$ emission intensity at 450 nm ($I_{450}$), and at 975 nm ($I_{975}$).

| Denote | Yb, at.% | Cr, at.% | $Cr^{3+}$ ($I_{450}$) | $Yb^{3+}$ ($I_{450}$) | $Yb^{3+}$ ($I_{975}$) |
|---|---|---|---|---|---|
| Yb0* | 0 | 0 | | | |
| Yb1* | 1 | 0 | | | 100 % |
| Yb0 | 0 | 0.5 | 100 % | | |
| Yb1 | 1 | 0.5 | 73 % | 100 % | 70 % |
| Yb5 | 5 | 0.5 | 29 % | 158 % | 170 % |
| Yb10 | 10 | 0.5 | 15 % | 84 % | 92 % |
| Yb20 | 20 | 0.5 | 4 % | 33 % | 48 % |
| Yb30 | 30 | 0.5 | 2 % | 17 % | 45 % |
| Yb50 | 50 | 0.5 | 0.7 % | 0.7 % | 7 % |
| Yb70 | 70 | 0.5 | 0.7 % | 1.5 % | 0.7 % |
| Yb100 | 100 | 0.5 | 0.6 % | 0.2 % | 5.1 % |

The lifetime of $Cr^{3+}$ ions decreases with increasing concentration of $Yb^{3+}$ ions. Fig. 6(a) shows the decay curves of R-line emission ($\lambda_{em}$ = 688 nm) under $\lambda_{exc}$ = 450 nm for Cr,Yb:YAG nanocrystals with Yb concentration from 1 at.% to 100%. PL lifetimes of $Cr^{3+}$ ions were calculated from the decay curves using a double exponential function:

$$I = I_0 + A_1 e^{(-t/\tau_1)} + A_2 e^{(-t/\tau_2)},$$

where I is the luminescence intensity at time t, and $\tau_1$ and $\tau_2$ are fast and slow decay components, respectively, $A_1$ and $A_2$ are constants. The dependence of lifetime on $Yb^{3+}$ concentration is shown in Fig. 6(b). The values of both components of $Cr^{3+}$ decay decrease with increasing $Yb^{3+}$ concentration. The lifetime of the slow decay component is in the range from 2.88(1) ms to 0.056(2) ms (black dots in Fig. 6(b)), and the lifetime of the fast component is from 0.58(2) ms to 0.0048(1) (red dots on Fig. 6(b)). The decay curve is a sum of $^2E_g \rightarrow ^4A_{2g}$ R-line and $^4T_{2g} \rightarrow ^4A_{2g}$ broadband transition decay. It was previously found that the radiative decay curves for R-lines are single exponential ones, while in the case of the broadband transitions, radiative decay curves consist of two components as in our case (see Fig. 6). The faster decay component is related to $Cr^{3+}$ ions in weak crystal field, probably, in dodecahedral sites [20].

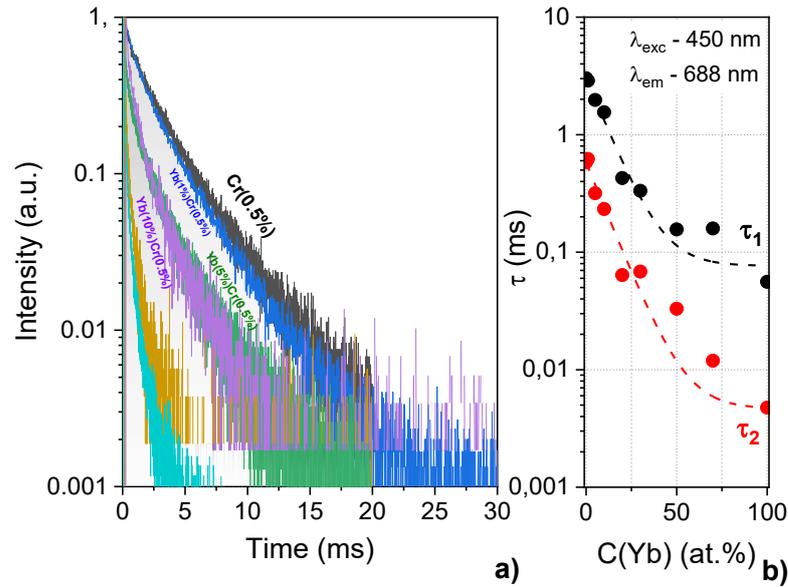

Fig. 6: Logarithmic plots of: a) $Cr^{3+}$ fluorescence decay curves and b) decay times versus $Yb^{3+}$ concentration in Cr,Yb:YAG nanocrystals monitored at $\lambda_{em}$ = 688 nm and measured at 450 nm at room temperature.

The concentration dependence of $Yb^{3+}$ emission shows a maximum at low doping content followed by drop to zero at high concentrations. Fig. 7(a) shows the photoluminescence spectra of $Yb^{3+}$ ions in Cr,Yb:YAG nanocrystals with $Yb^{3+}$ content in the range from 1 at.% to 100 at.% at $\lambda_{exc}$ = 450 nm. Room temperature luminescence spectra of $Yb^{3+}$ ions correspond to the intermanifold $^2F_{7/2} \leftrightarrow {}^2F_{5/2}$ transitions of $Yb^{3+}$ ions and are characterized by sharp and narrow emission bands with three most intense at 940 nm, 967 nm, and 1028 nm. The spectra of $Yb^{3+}$ emission at $\lambda_{exc}$ = 450 nm or 975 nm show an increase in emission intensity with an increase in the concentration of $Yb^{3+}$ ions up to 5 at.% with a subsequent drop (Fig. 7(b)). It should be noted that the concentration dependence of the $Yb^{3+}$ emission intensity exhibits the same pattern for both $\lambda_{exc}$ = 450 nm and 975 nm.

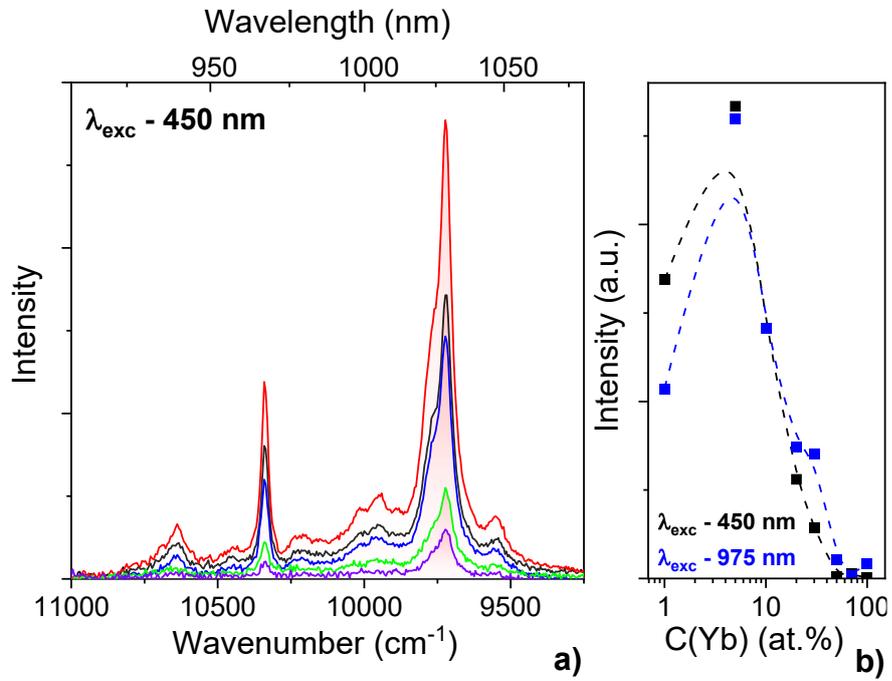

Fig. 7. a) Photoluminescence spectra of $Yb^{3+}$ ions. b) Dependence of luminescence intensity on the $Yb^{3+}$ concentration in Cr,Yb:YAG nanocrystals measured at 450 nm (black line) and 975 nm (blue line) at room temperature.

The measured lifetime of $Yb^{3+}$ ions decreases with increasing $Yb^{3+}$ concentration. Fig. 8(a) shows the fluorescence decay curves of $^2F_{7/2} \leftrightarrow {}^2F_{5/2}$ transitions of $Yb^{3+}$ ions ($\lambda_{em}$ = 1028 nm) at $\lambda_{exc}$ = 450 nm for Cr,Yb:YAG nanocrystals with Yb concentrations from 1 at.% to 100% at.%. PL lifetime of $Yb^{3+}$ ions was calculated from decay curves using a single exponential function:

$$I = I_0 + A_1 e^{(-t/\tau)}$$

where I is the luminescence intensity at time t, $\tau$ is lifetime, $A_1$ is constant. The concentration dependence of lifetime of $Yb^{3+}$ ions is shown in Fig. 6(b) for both $\lambda_{exc}$ = 450 nm and 975 nm. The measured lifetimes of $Yb^{3+}$ ions decrease with increasing $Yb^{3+}$ concentration. The lifetimes were in the range from ~ 2.5 ms to ~ 0.002 ms being the same for both $\lambda_{exc}$ = 450 nm and 975 nm (Fig. 8(b)). The measured lifetimes are collected in the Table 4.

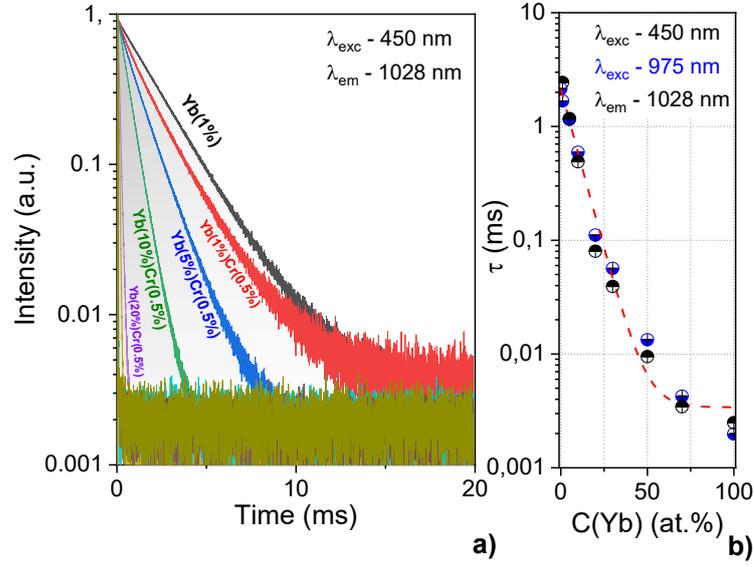

Fig. 8: Logarithmic plots of: a) $Yb^{3+}$ fluorescence decay curves and b) decay time versus $Yb^{3+}$ concentration in Cr,Yb:YAG nanocrystals monitored at $\lambda_{em}$ = 1028 nm with $\lambda_{exc}$ = 450 nm (black dots) and 975 nm (blue dots) at room temperature.

The efficiency of energy transfer between $Cr^{3+}$ and $Yb^{3+}$ ions increased with increasing $Yb^{3+}$ concentration. The energy transfer efficiency was calculated according to the equation:

$$\eta = 1 - \frac{Cr^{3+}(\tau_1)}{Cr^{3+}(\tau_1^0)}$$

where η is the efficiency of energy transfer, $Cr^{3+}(\tau_1)$ is the lifetime of $Cr^{3+}$ ions (slow component) in the presence of $Yb^{3+}$, and $Cr^{3+}(\tau_1^0)$ is the lifetime of $Cr^{3+}$ ions in the absence of $Yb^{3+}$. The measured data are collected in the Table 4. It can be seen that increasing $Yb^{3+}$ concentration caused an increase in the energy transfer probability from $Cr^{3+}$ ions to $Yb^{3+}$ ones.

Table 4: Measured lifetimes of $Cr^{3+}$ and $Yb^{3+}$ ions in Cr,Yb:YAG nanocrystals: lifetimes of $Cr^{3+}$ fast ($\tau_1$) and slow ($\tau_2$) components at $\lambda_{exc}$ = 450 nm, and of $Yb^{3+}$ ions at $\lambda_{exc}$ = 450 nm ($\tau_{450}$) and $\lambda_{exc}$ = 975 nm ($\tau_{975}$).

| Denote | Yb, at.% | Cr, at.% | $Cr^{3+}$ ($\tau_1$), ms | $Cr^{3+}$ ($\tau_2$), ms | $Yb^{3+}$ ($\tau_{450}$), ms | $Yb^{3+}$ ($\tau_{975}$), ms | η |
|---|---|---|---|---|---|---|---|
| Yb0* | 0 | 0 | | | | | |
| Yb1* | 1 | 0 | | | | 2.1376(5) | |
| Yb0 | 0 | 0.5 | 3.02(2) | 0.58(2) | | | |
| Yb1 | 1 | 0.5 | 2.88(1) | 0.62(1) | 2.43(1) | 1.676(2) | 0.05 |
| Yb5 | 5 | 0.5 | 1.98(1) | 0.32(3) | 1.17(1) | 1.1521(6) | 0.34 |
| Yb10 | 10 | 0.5 | 1.55(2) | 0.233(4) | 0.489(1) | 0.5926(1) | 0.49 |
| Yb20 | 20 | 0.5 | 0.43(1) | 0.064(1) | 0.0800(1) | 0.1107(2) | 0.86 |
| Yb30 | 30 | 0.5 | 0.33(1) | 0.068(1) | 0.0392(1) | 0.0566(2) | 0.89 |
| Yb50 | 50 | 0.5 | 0.16(1) | 0.033(1) | (0.00949(4) | 0.0133(1) | 0.95 |
| Yb70 | 70 | 0.5 | 0.16(1) | 0.0119(1) | 0.00344(2) | 0.0043(1) | 0.95 |
| Yb100 | 100 | 0.5 | 0.056(2) | 0.0048(1) | 0.00249(1) | 0.00198(3) | 0.98 |

### 3.3 LIWE properties

The shape of the LIWE spectrum was similar to the one previously found for other nanocrystals and different from the spectrum obtained for Yb-doped transparent ceramics. Excitation of Cr,Yb:YAG nanocrystals by focused 975 nm laser beam in vacuum caused an appearance of white light emission within an entire visible and near infrared region [21]. Similar to our previous studies [1,5,6,8,22,23], the measured LIWE spectra were not corrected for camera sensitivity. More detailed explanation can be found in [2]. The shape of the emission spectrum is similar to one previously found for Yb:YAG nanocrystals, including the appearance of an extra peak centered at the excitation wavelength [8]. The recorded shape of the emission spectrum differs from the one obtained in Yb-free compounds [23]. It seems that the additional peak in the LIWE spectra does not correspond to $^2F_{7/2} \leftrightarrow ^2F_{5/2}$ transitions of $Yb^{3+}$ ions, since this emission disappears during generation of white light emission [8]. Moreover, our earlier studies have shown that fully concentrated $Yb_2O_3$ transparent ceramics exhibit no LIWE peak at excitation wavelength [1]. In fact, $Yb_2O_3$ transparent ceramics show a decrease in emission intensity above 800 nm that can be caused by $Yb^{3+}$ absorption. The origin of the difference in the emission spectra of different Yb-doped phosphors remains unclear.

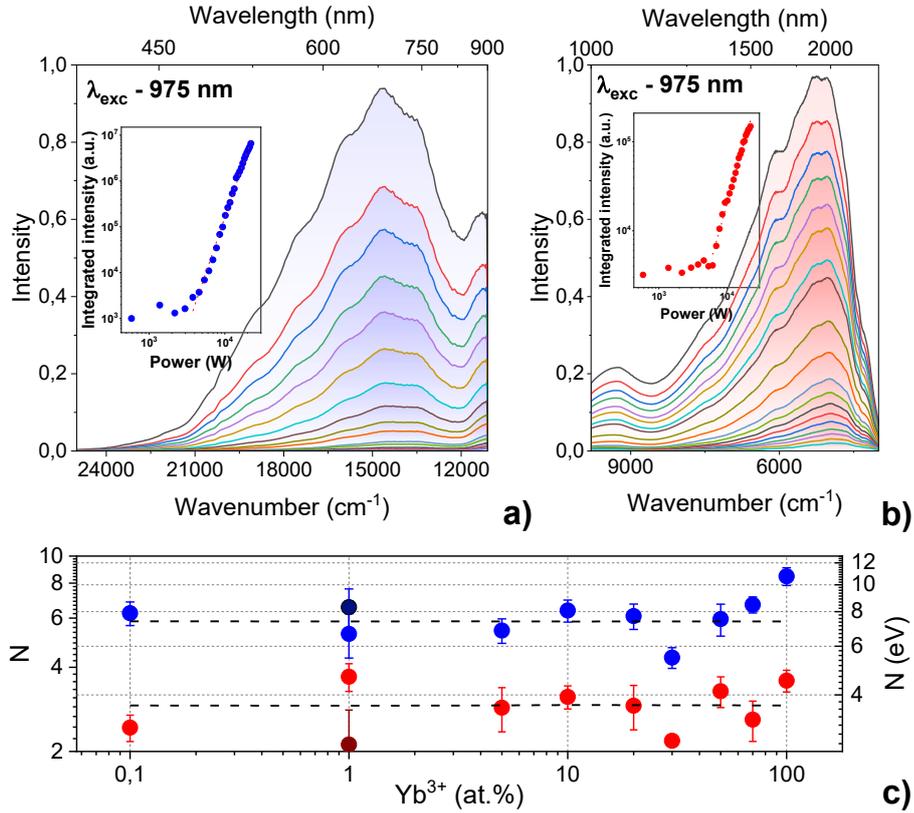

Fig. 9. a) Anti-Stokes and b) Stokes parts of LIWE spectra of Cr,Yb:YAG nanocrystals at $\lambda_{exc}$ =975 nm in vacuum. The inset shows a logarithmic plot of LIWE power dependence for Yb50 sample. c) Logarithmic plot of N parameter versus the concentration of $Yb^{3+}$ ions (starting point at 0.1 represents Cr-only doped sample).

A change in the concentration of $Yb^{3+}$ ions does not affect the number of phonons required for the generation of photoelectrons, as well as the LIWE threshold. One of the prominent features of LIWE is the exponential increase in LIWE intensity after exceeding the threshold. The threshold values for the anti-Stokes and Stokes parts were found to be $4(1)\cdot10^3$ W/cm² and $5(2)\cdot10^3$ W/cm², respectively. It should be noted that the obtained threshold is higher than for Yb:YAG nanocrystals ($3.5\cdot10^3$ W/cm² and $2\cdot10^3$ W/cm² for anti-Stokes and Stokes parts, respectively) [8]. Remarkably, the power threshold for the anti-Stokes part of LIWE spectrum was higher than for the Stokes part, that is rare for white light emission [8,23]. Schematically, the power dependence can be described as:

$$I_{LIWE} \propto P^N,$$

where $I_{LIWE}$, P, and N are LIWE intensity, excitation power, and the minimal number of absorbed photons required for ionization, respectively. Fig. 9(c) shows the power dependence of N parameter for Cr,Yb:YAG nanocrystals. A change of $Yb^{3+}$ concentration does not affect the number of absorbed photons required for electron emission (N parameter). The detected value of the N parameter was from 4 to 7 with a mean of 5.8, and from 2 to 4 with a mean of 2.9 for the anti-Stokes and Stokes parts of the spectrum, respectively.

The measured rise and decay times of LIWE were in the range of seconds without any concentration-dependent pattern. It was shown that turning on/off the excitation beam caused a steady increase/decrease in the emission intensity. The temporal evolution of the emission intensity can be fitted by a single exponential function similar to those described above for the luminescence lifetimes. It was determined that there is no regularities in the values of rise and decay times of LIWE for Cr,Yb:YAG nanocrystals. For example, Fig. S2 shows the temporal evolution of LIWE intensity after turning on/off laser excitation for Yb1 sample. The measured rise and decay times were 3-10 s and 0.1-0.5 s, respectively. A similar pattern was found earlier for Yb:YAG nanocrystals: the rise and decay times were 4(2) s and 0.3(2) s, respectively [8]. A higher value of rise time than that of decay time is a general pattern for LIWE [8,22]. It should be noted that the measured LIWE decay and rise times for materials with low thermal conductivity (such as nanocrystals) were an order of magnitude shorter than for materials with high thermal conductivity. For example, the time response of LIWE for graphene foam (0.5–0.9 ms) [24], tungsten filament (4–7 ms) [25], and Cr:YAG transparent ceramics (6-9 ms) is much shorter than the times reported here.

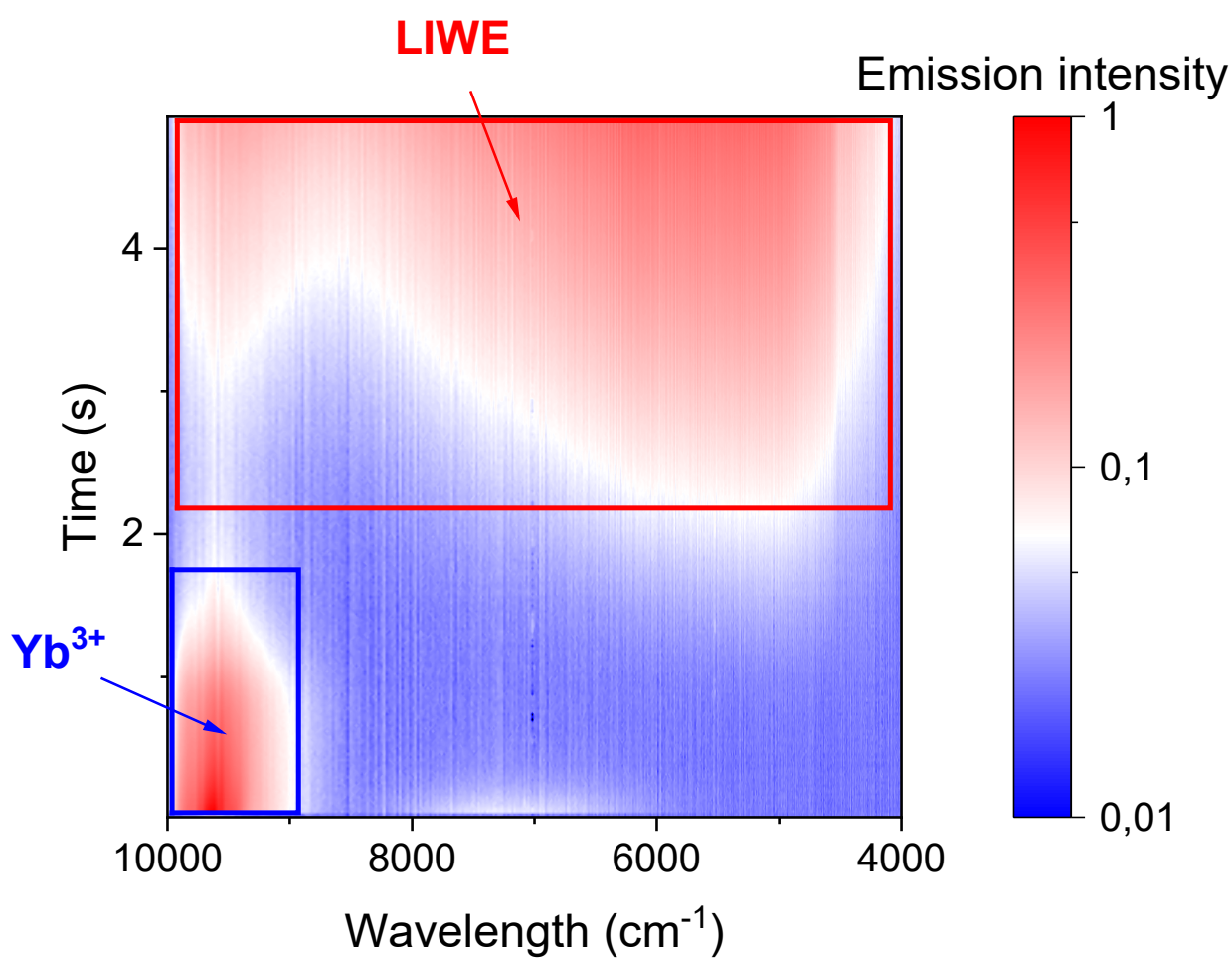

Fig. 10: Time map of LIWE evolution of Yb30 sample under 975 nm excitation (Stokes part)

Characteristic feature of Yb-doped materials is a decrease in the $Yb^{3+}$ emission with LIWE increase. Fig. 10 shows the time evolution of the Stokes part of LIWE spectra. At first, the strong emission band centered at 1030 nm is observed corresponding to $^2F_{5/2} \rightarrow {}^2F_{7/2}$ electronic transition of $Yb^{3+}$ ions. Laser irradiation leads to an increase of LIWE intensity with a simultaneous decrease of $Yb^{3+}$ intensity (Fig. S3). $Yb^{3+}$ emission disappears after 2 s. Earlier the similar effect was detected for Yb:YAG nanocrystals. In absence of LIWE, the decrease in the $Yb^{3+}$ emission intensity during an excitation of the sample in vacuum was 10% that was caused by the heating of the sample, see our earlier paper [8]. However, when LIWE appears, the $Cr^{3+}$ disappears completely which can be caused by ionization process when absorbed energy of $Yb^{3+}$ ions was carried out by ejected electrons.

A similar effect of appearance and decrease of $Cr^{3+}$ emission was found for the anti-Stokes part of the emission spectra. Fig. 11 shows the temporal evolution of LIWE spectra for Cr,Yb:YAG nanocrystals. Turning on the laser excitation caused the appearance of $^4T_{2g} \rightarrow {}^4A_{2g}$ broadband emission of $Cr^{3+}$ ions. It should be noted that the $Cr^{3+}$ emission appears after certain time and increases during 3 seconds until reaching maximum. An increase of the excitation time causes a decrease of $Cr^{3+}$ emission intensity with a simultaneous increase of LIWE intensity, as for $Yb^{3+}$ emission (Fig. 10). At the low energy edge of the time map, one can observe a part of the excitation beam. LIWE increase leads to decrease of the reflected beam intensity indicating an increase in absorption (Fig. S4), similar the one previously reported for Yb:YAG nanocrystals [8]. $Cr^{3+}$ emission is probably caused by energy transfer from $Yb^{3+}$ to $Cr^{3+}$ ions, but other explanations also can be proposed. A specific feature of the spectra is the presence of $^4T_{2g} \rightarrow {}^4A_{2g}$ emission only without any traces of R-lines. Such emission was observed earlier for $Cr^{3+}$:YAG single crystals at 975 nm excitation [26]. It is important to note that $Cr^{3+}$ upconversion emission increases with increasing sample temperature. Increase/decrease of the $Cr^{3+}$ emission intensity (Fig. 11) can be caused by an increase of the sample temperature leading to an increase of $Cr^{3+}$ upconversion emission.

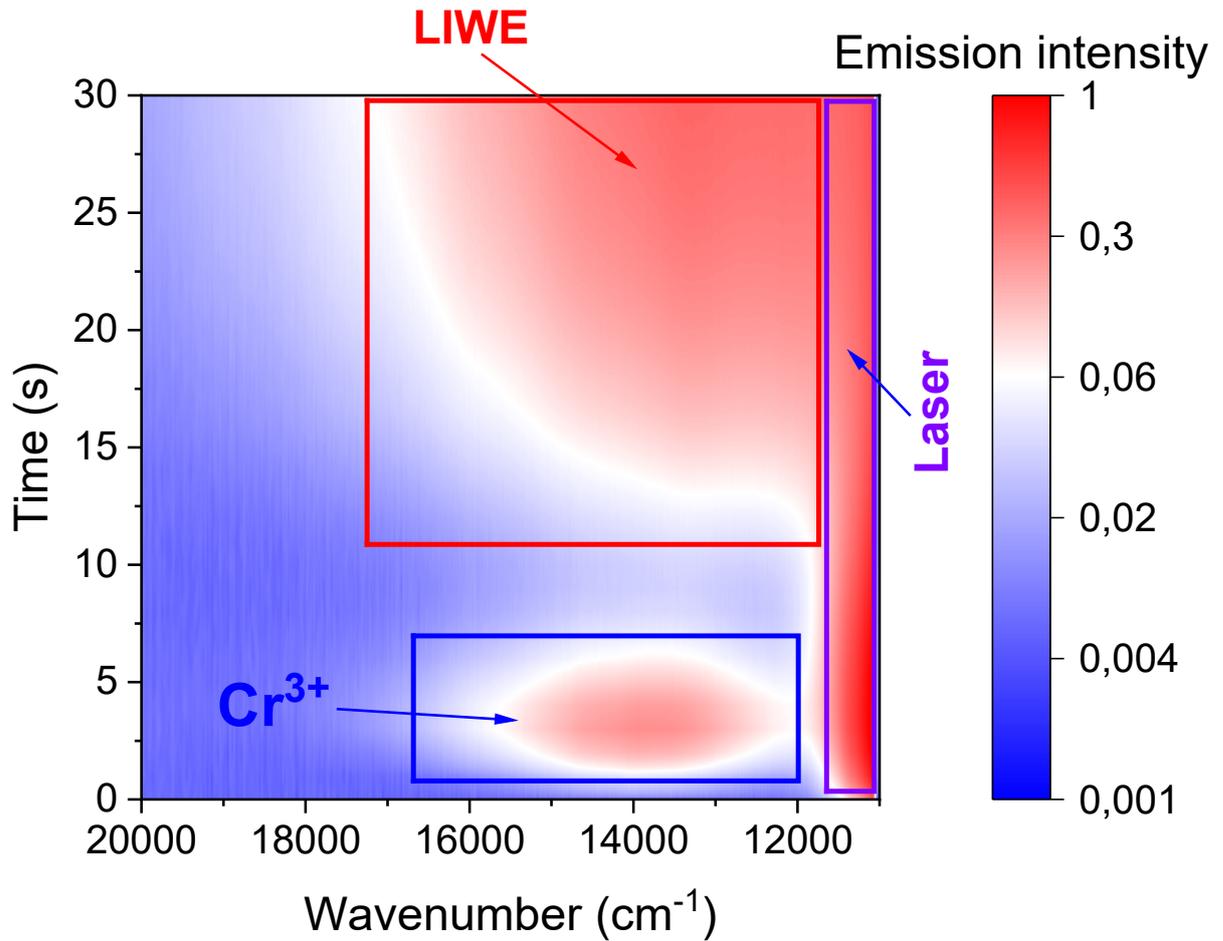

Fig. 11: Time map of LIWE evolution in Yb1 sample under 975 nm excitation (anti-Stokes part)

## 4. Discussion

On the basis of the similarity in the power dependence of emission, multiphoton ionization model was proposed to explain the features of LIWE. It is difficult to understand the results of this work without reference to the current model of LIWE phenomenon. At present, several different models have been proposed to explain LIWE phenomenon including photon avalanche, thermal avalanche, electron-hole recombination, etc. We suggest that a multiphoton ionization model can be proposed to describe the nature of LIWE phenomenon. Beforehand, multiphoton ionization model was proposed to describe the interaction of high-density pulsed laser beam with atoms, which caused the appearance of free electrons [27]. The main difference is that in our case continuous excitation with a density of $10^2$-$10^5$ W/cm$^2$ was used, while multiphoton ionization occurs under pulsed laser excitation with a density of $10^{10}$-$10^{20}$ W/cm$^2$ [1]. The applicability of the multiphoton ionization model is based on the similarity of the properties of white light emission and multiphoton ionization as the exponential growth, threshold dependence, and saturation are common to both cases. Moreover, the appearance of white light emission caused a decrease in the resistivity of irradiated materials by four orders of magnitude indicating the formation of charge carriers (for example, in $Yb_2O_3$ transparent ceramics [1]).

Multiphoton ionization causes the emission of photoelectrons, which then return to the surface, recombine and, as a result, emit white light. The multiphoton ionization model can be a skeleton on which other branches can be placed to build a correct model. The multiphoton ionization model cannot explain the differences in LIWE rise and decay times in different compounds. In addition, this model cannot explain the presence of such long rise/decay times because LIWE must appear/disappear immediately after the excitation beam is turned on/off. For example, the calculated radiative transition rate for LIWE was 2-10 $s^{-1}$ for our materials, and up to $10^3$ $s^{-1}$ in some other cases [22,24], while ionization rates for multiphoton ionization of xenon were found to be more than $10^{13}$ $s^{-1}$ [28]. It should be noted that the multiphoton ionization rate was measured by recording the emitted electrons, while in our case the emission of photons was recorded. We assume that the white light emission is a byproduct of electron emission, as the ejected electrons return to the surface and lose energy by radiative recombination process that is accompanied by light [1].

The exponential growth of the emission intensity indicates an electron avalanche process taking place in $Yb^{2+}/Yb^{3+}$ mixed valence pairs. When the excitation energy is below the power threshold, no ionization is observed. After exceeding the threshold, the first electron is released providing the next ionization. The first ejected electron leads to the appearance of several subsequent photoelectrons, which, in turn, favor further ionization causing an avalanche-like process [29]. The previously detected increase in laser absorption during the onset of LIWE [8] is an indication that the avalanche ionization process is involved in LIWE. One of the possible explanation of the avalanche-ionization process can be the formation of $Yb^{2+}/Yb^{3+}$ mixed valence pairs during LIWE, more detailed description can be found in our earlier papers [5,8,22]. Ionization of one $Yb^{2+}/Yb^{3+}$ pair favors the formation of additional $Yb^{2+}/Yb^{3+}$ pairs, which form four mixed valence pair in the next ionization cycle, and so one. The appearance of $Yb^{2+}$ ions after LIWE generation [8] is consistent with this theory.

$Yb^{3+}$ ions play an important role in the LIWE generation, as they acts as the main laser beam absorbing centers. Therefore, the presence of $Cr^{3+}$ ions should affect the LIWE properties. An increase in the concentration of $Yb^{3+}$ ions in Cr,Yb:YAG nanocrystals leads to an increase in the probability of energy transfer between $Cr^{3+}$ and $Yb^{3+}$ ions (Table 4). So, the minimal number of absorbed photons required for ionization (N parameter) should increase with increasing $Yb^{3+}$ concentration as a most part of the absorbed energy is lost due to the energy transfer to the chromium ions. However, our research shows that N parameter does not depend on the concentration of $Yb^{3+}$ ions. Moreover, the decrease in the lifetime of $Yb^{3+}$ ions (Fig. 8) also does not affect the LIWE parameters.

The independence of N on the spectroscopic properties of $Yb^{3+}$ ions indicates that the ionization rate for multiphoton ionization of $Yb^{2+}/Yb^{3+}$ mixed valence pairs in Cr,Yb:YAG nanocrystals is much faster than the radiative transition rate for white light emission. Taking into account the lifetimes of $Yb^{3+}$ ions (Table 4), we can conclude that the ionization rate for multiphoton ionization of Cr,Yb:YAG nanocrystals is higher than $10^5$ s$^{-1}$. The measured decrease in $Yb^{3+}$ and $Cr^{3+}$ emission intensities during LIWE generation is consistent with the proposed explanation (Fig. 10, Fig. 11). The energy accumulated in $Yb^{3+}$ ions participates more likely in the ionization process than in $^2F_{5/2} \rightarrow {}^2F_{7/2}$ electronic transition of $Yb^{3+}$ ions. A similar result was previously found for Yb:YAG nanocrystals [8], where the emission of $Yb^{3+}$ ions disappears only after the start of LIWE.

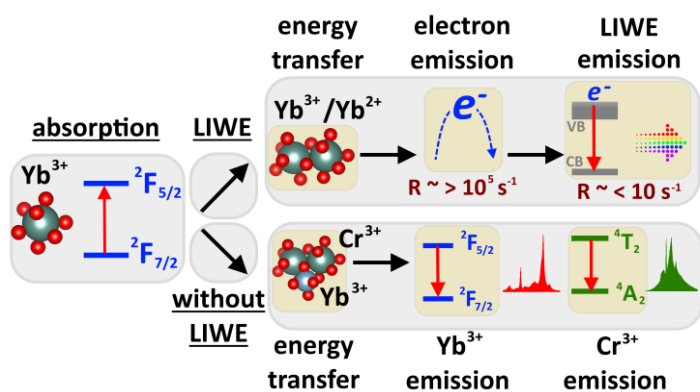

Fig. 12: Schematic representation of the luminescence of Cr,Yb:YAG nanocrystals upon excitation in vacuum with and without LIWE generation.

Taking into account that the white light emission is visible for several seconds after switching off the laser excitation, it can be concluded that the generation of photons occurs after the ejection of electrons which is followed by their recombination. Most likely, electrons can be trapped for a few seconds after ionization providing a decrease in resistivity during LIWE generation [1]. White light emission occurs during the recombination of trapped electrons with the ionizing center ($Yb^{2+}/Yb^{3+}$ mixed valence pairs in the case of Cr,Yb:YAG nanocrystals). Fig 12 shows a schematic illustration of emission mechanism of Cr,Yb:YAG nanocrystals under excitation in vacuum with and without LIWE generation. Proposed explanation is general and does not include specific details such as the type of electron traps. Moreover, de-trapping of electrons occurs faster at high temperatures, while the opposite behavior is found for LIWE phenomenon. An increase of the excitation power (as a result of the temperature increase) provides longer decay times of LIWE [24].

Despite the success in applying multiphoton ionization theory to LIWE, there is still a discrepancy between experimental data and theory. So, the proposed explanation is far from perfect and more studies are needed. The further work should therefore include the effect of defects on LIWE properties.

## 5. Conclusions

Synthesized Cr,Yb:YAG nanocrystals are characterized by pure YAG crystal structure with an average grain size of ~ 30 nm. Cr,Yb:YAG nanocrystals were synthetized by Pechini method. The microstructure of the samples was characterized by TEM and XRD techniques. X-ray diffraction has shown an absence of any additional phases and linear decrease of the lattice parameter from 11.9 Å to 12.3 Å with increasing $Yb^{3+}$ concentration. The TEM study has revealed the fine crystal structure of synthesized Cr,Yb:YAG nanocrystals.

An increase in the $Yb^{3+}$ concentration causes an increase in the efficiency of energy transfer affecting the optical properties of Cr,Yb:YAG nanocrystals. Diffuse reflectance spectra of Cr,Yb:YAG nanocrystals show the presence of absorption bands corresponding to $Yb^{3+}$, $Cr^{3+}$, $Cr^{6+}$ ions, and color centers. $Cr^{3+}$ ions cause an increase in absorption in the UV region due to the appearance of strong additional absorption bands centered at 270 nm and 370 nm. The excitation spectra of $Yb^{3+}$ ions followed the same pattern as the spectra of $Cr^{3+}$ ions indicating $Cr^{3+} \rightarrow Yb^{3+}$ energy transfer. The $Cr^{3+}$ photoluminescence intensity gradually decreased with increasing $Yb^{3+}$ concentrations, as well as $Yb^{3+}$ emission itself. The lifetimes of $Cr^{3+}$ and $Yb^{3+}$ ions decreased three orders of magnitude with increasing $Yb^{3+}$ concentration. The energy transfer efficiency between $Cr^{3+}$ and $Yb^{3+}$ ions increased with increasing $Yb^{3+}$ concentration and was in the range from 0.05 to 0.98.

Increasing energy transfer efficiency did not affect the LIWE properties. Excitation of the Cr,Yb:YAG nanocrystals with a focused 975 nm laser beam in vacuum caused the appearance of white light emission covering the entire visible and near infrared region. It was found that the change in the concentration of $Yb^{3+}$ ions does not affect the number of absorbed photons required for emission of electron (N parameter). The detected value of N parameter ranged from 4 to 7 with a mean of 5.8 and from 2 to 4 with a mean of 2.9 for anti-Stokes and Stokes parts, respectively. There was no regularity in the rise and decay times of LIWE (3-10 s and 0.1-0.5 s, respectively).

The independence of LIWE properties on $Yb^{3+}$ concentration is explained by fast radiative recombination during multiphoton ionization as compared to white light emission. The multiphoton ionization model was used to explain LIWE properties in the Cr,Yb:YAG nanocrystals. Based on the calculated difference in radiative recombination rates between LIWE and multiphoton ionization processes, it was concluded that white light emissions occurs after electron emission. We suppose that the white light emission is byproduct of electron emission, when the emitted electrons return to the surface losing energy in a radiative recombination process and producing light. This paper shows that N parameter does not depend on the concentration of $Yb^{3+}$ ions. Moreover, the decrease in the lifetime of $Yb^{3+}$ ions also does not affect the LIWE parameters. The independence of the N parameter on the spectroscopic properties of $Yb^{3+}$ ions indicates that the ionization rate for multiphoton ionization of $Yb^{2+}/Yb^{3+}$ mixed valence pairs in Cr,Yb:YAG nanocrystals is much faster than the radiative transition rate of white light emission. Taking into account the measured lifetimes for $Yb^{3+}$ ions, one can conclude that the ionization rate for multiphoton ionization is higher than $10^5\,s^{-1}$. Taking into account that the white light emission is visible for several seconds after switching off the laser excitation, it can be concluded that photon emission occurs after electron emission. Most likely, after ionization, the electrons can be trapped for a few seconds.

## Acknowledgement

This work was supported by Polish National Science Centre, grant: PRELUDIUM-18 2019/35/N/ST3/01018.